\RequirePackage{lineno}

\documentclass[12pt]{iopart}
\usepackage{epsfig}

\begin{document}

\title{The Yale liquid argon time projection chamber}

\author{Alessandro Curioni, Bonnie T. Fleming, Mitchell Soderberg}

\address{Yale University,
Physics Department,
217 Prospect St.
New Haven CT 06511 - USA}
\ead{alessandro.curioni@yale.edu}

\begin{abstract}
In this paper we give a thorough description of a liquid argon
time projection chamber designed, built and operated at Yale.
We present results from a calibration run where cosmic rays have been
observed in the detector, a first in the US.
\end{abstract}

\maketitle

\section{Introduction}

Liquid argon time projection chambers (LAr TPC) are nearly optimal
detectors for neutrino experiments looking for $\nu _e$ appearance in a
$\nu_{\mu}$ beam in the energy range 0.5 -- 5 GeV. The LAr TPC
technology has been proposed for the measurement of $\theta _{13}$, CP
violation in the neutrino sector and  determination of the mass
hierarchy (e.g.~\cite{proposal1, proposal2, proposal3}), and to study
the MiniBooNE low energy anomaly \cite{miniboone, microboone}.  
The technique is equally promising for proton decay searches and
astrophysical applications \cite{bueno}.   
A LAr TPC for neutrino physics was first proposed in 1977 by Carlo
Rubbia \cite{rubbia}. The ICARUS collaboration established  a vigorous
R\&D program, which produced decisive steps in defining the technology
and its applicability to particle physics (e.g.~\cite{3ton, 50l,
t600}).    

Images taken in a LAr TPC are comparable in quality with pictures from
bubble chambers. 
As it is the case in bubble chambers, events can be analyzed by
reconstructing 3-momentum and particle type for each track in the
event image, with a lower energy threshold of few MeV for electrons
and few tens of MeV for protons.
The particle type can be determined from measuring the energy loss
along the track ($dE/dx$) or from topology (i.e. observing the decay 
products).   
The calorimetric performance of a LAr TPC ranges from good to
excellent, depending on event energy and topology; for details, see
the very descriptive experimental paper \cite{t600}, and references
therein.   

Liquid argon has a density of 1.4 g/cm$^3$, is fairly inexpensive
($\sim$1~USD/liter) and readily available in large amounts. 
This makes LAr an attractive option as active medium for very massive
detectors (several tens of ktons), as required by most applications in
contemporary neutrino physics.   
The TPC technology offers a practical way to image volumes as large as
several thousands of cubic meters.

A staged program toward a very massive (50-100 kton) LAr TPC for
neutrino physics is ongoing in the United States, following the
recommendations of the Neutrino Scientific Assessment Group
(\verb+NuSAG+) in 2007 \cite{nusag}.    
As part of this larger, comprehensive US effort, a LAr TPC has
been developed at Yale starting in 2005.  
The detector was commissioned in early 2007, and cosmic ray
tracks were imaged. This is the first LAr TPC prototype developed in
the US to image tracks.    

In this paper we give a detailed description of the experimental
apparatus (Sec.~2), present results from the calibration runs
(Sec.~3), and comment on foreseen developments (Sec.~4). 

\section{Description of the detector}

A LAr TPC is a position sensitive liquid ionization chamber. When 
the LAr in the active volume is ionized by a charged particle, a
number of free electrons, equal (on average) to the energy lost by the 
particle divided by the W-value, is produced in LAr. The W-value in LAr
is 23.6~eV \cite{Wvalue}. The ionization electrons drift under the
action of a uniform electric field over the distance between the
interaction point and the readout electrodes. In this TPC the maximum
distance is 16 cm; over this distance, in a dense medium as LAr, the
lateral spread of the drifting charge due to diffusion is negligible.  
The drifting charge induces a signal on the readout electrodes, two
parallel planes of wires in our case. 
Free electrons can attach to electronegative impurities and be lost in
the drift region, drastically reducing the size of the induced
signal. For this reason the concentration of electronegative
impurities in the drift region has to be less than one part per
billion.  
A first signal is induced on the wires immediately facing the drift
region ({\it induction plane}).In this detector this wire plane also
acts as a Frisch grid for the second plane ({\it collection
plane}). The electric field between the two wire planes is at least
twice the one in the drift region in order to have good transparency
through the induction plane for the drifting charge. Only when the
drifting electrons are past the induction plane is a signal induced on
the collection plane. The field lines terminate on the collection
wires, where the charge is collected. The wire number
defines one spatial coordinate, therefore two wire planes provide two
spatial coordinates, and the third spatial coordinate is measured by
the drift time. The energy deposition is measured from the amplitude
of the induced signal. 

\begin{figure}[htb]
  \includegraphics[width=.5\textwidth]{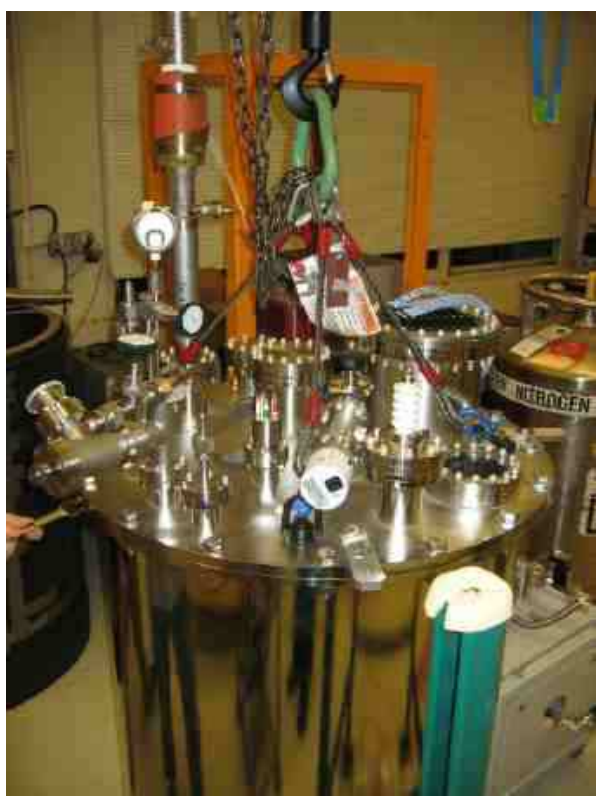}
  \includegraphics[width=.5\textwidth]{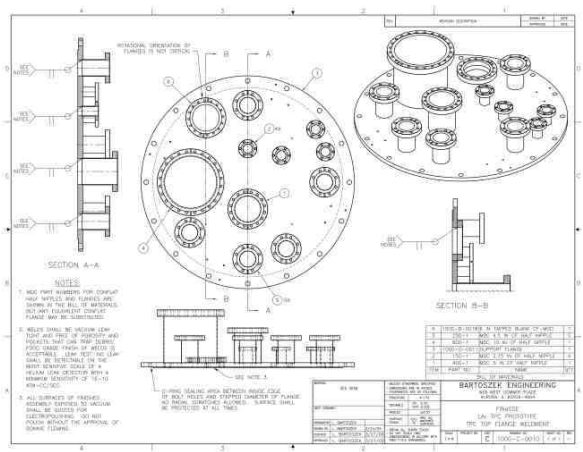}
 \caption{{\it Left:} Picture of the top flange, described in the
 text. {\it Right:} Top flange, technical drawing.} 
\label{f1a}
\end{figure}

\begin{figure}
  \centering
  \includegraphics[width=.75\textwidth]{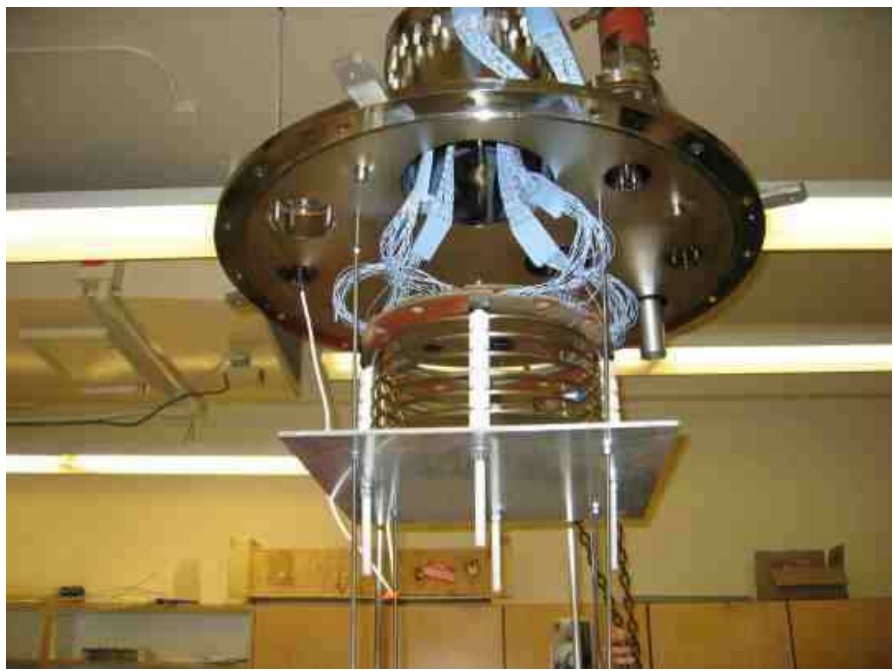}
  \caption{The TPC, fully cabled, hanging from the top flange. }
\label{f1b}
\end{figure}

\subsection{Cryogenics}

The LAr TPC is housed in a cylindrical stainless steel vessel, 111~cm
long with a diameter of 72.4~cm, for a total volume of about 460~l. 
The inner surface of the vessel has been electropolished. 
Prior to filling with ultra pure LAr, the vessel is evacuated to few
10$^{-6}$ mbar at LAr temperature (87 K). The vessel is cooled to LAr
temperature by an open bath filled with commercial, non-purified LAr.  
It takes about five hours to cool down the system and fill the inner
vessel up to the level where the TPC structure is fully covered by
LAr, equivalent to $\sim$250~l of ultra-pure LAr. The total LAr
consumption is $\sim$ 1,000~l for a 24 hr long experiment, about half
of it going directly to the open bath, and half through the
purification system to the inner vessel. 

The top flange of the TPC vessel (Fig.~\ref{f1a}) houses several
ports: 
feedthroughs for the TPC high-voltage, signal cables, test pulse and
capacitive level meter, high voltage feedthrough and optical
feedthrough for a purity monitor mounted inside the vessel, pumping
line, filling line, relief valve, pressure gauges, and a window. 
All the seals on the top flange are either CF or VCR. The top flange
itself is sealed using a Viton O-ring, therefore the system is not
designed to be vacuum-tight at LAr temperature in steady state. The
adopted filling procedure is to break the vacuum using ultra-pure cold
Ar gas when the outer LAr open bath is about half full. Once in steady 
state the system runs at an overpressure of 0.3 atm \footnote{The
relief valve has been provided by H. Jostlein of FNAL.}.   

\subsection{TPC and Electronics}

\begin{figure}
  \centering
  \includegraphics[width=\textwidth]{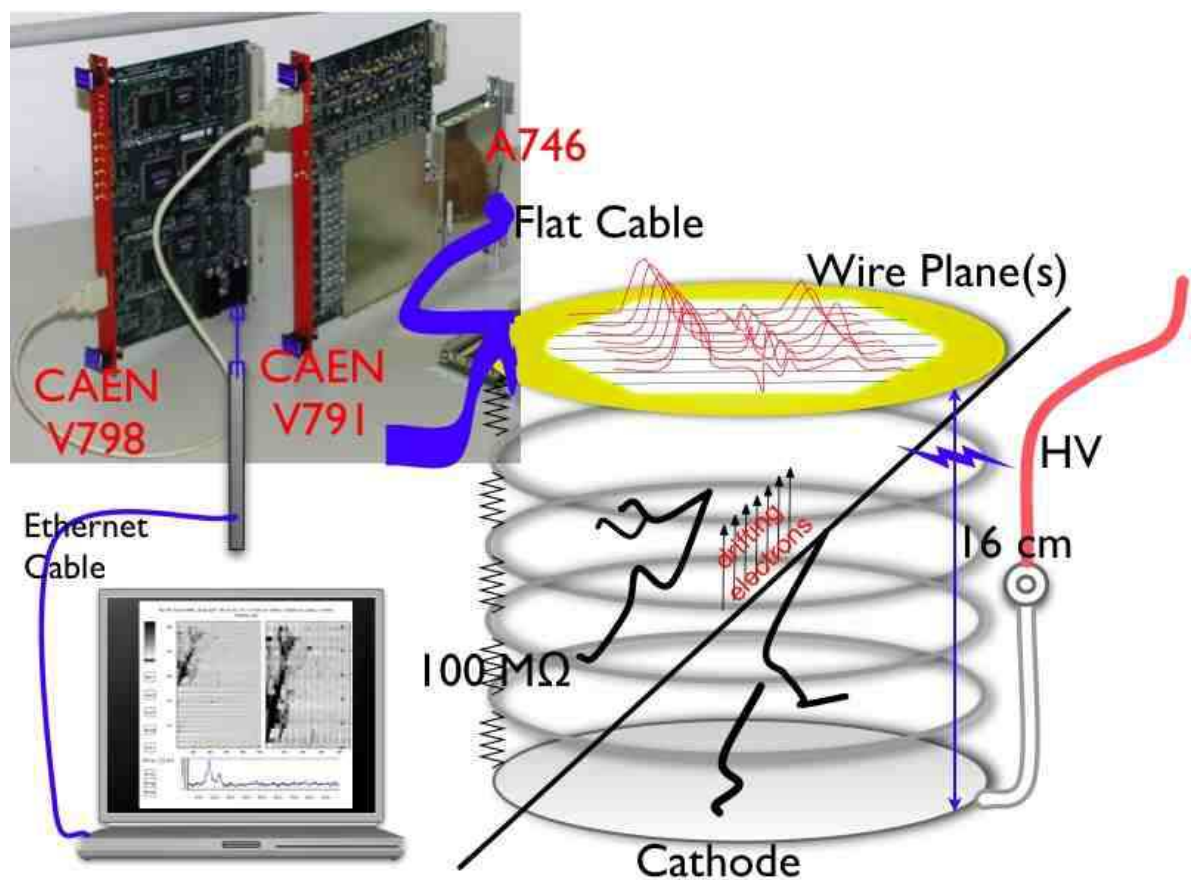}
  \caption{Schematic of the TPC readout and electronic chain,
  not-to-scale. (See text.)} 
\label{schematics}
\end{figure}
The instrumented volume of the LAr TPC (Fig.~\ref{f1b}) is a
cylinder of 33~cm diameter and 16~cm tall. A schematic of the TPC is 
shown in Fig.~\ref{schematics}. The field cage is made of 6    
hollow stainless steel rings (6~mm diameter),
separated by Teflon spacers, and electrically connected through a
chain of 100~M$\Omega$ resistors.  
There are two parallel readout wire planes of hexagonal shape with
smooth edges. Each plane has  50 wires with a wire pitch of 5
mm. The longest wire is 26 cm long, the shortest 13 cm. The wires
are soldered on a G-10 frame, 6~mm thick with a layer of etched copper
on one side.  
The first wire plane, directly facing the drift region, acts as an
induction plane, with the collection plane 6~mm behind it.  
The collection wires run at a 60 degree angle with respect to the
induction wires. 
The induction plane was intended as a Frisch grid for the collection
plane, and little care has been taken in decoupling it from the
high-voltage chain. Nonetheless, it was connected to the readout
chain.  
Flat cables are soldered to the wire planes and reach the readout
electronics through a signal feedthrough \footnote{From INFN-Padova /
  ICARUS.}, which holds 512 channels. 
The readout electronics and data acquisition (DAQ) have been provided
by the INFN-Padova / ICARUS group. A detailed description of these can
be found in \cite{t600}. To summarize, the front-end electronics
consists 
of three main modules: 
\begin{enumerate}
\item a decoupling board (\verb+A746+) mounted on the back side of a
  VME-like crate, which receives 32 analog signals and passes them to
  the analog board;
\item an analog board (\verb+CAEN V791+), housed in the VME-like
  crate, with 32 amplifiers, which also performs multiplexing and the
  analog-to-digital conversion (10 bits, 40 MHz). The front-end
  preamplifiers have a short time constant (1.6 $\mu$s) and behaves as
  an approximate differentiator;
\item a digital board (\verb+CAEN V798+), in a separate VME
  crate. Each pair of analog/digital boards is connected through a
  serial link, and the data are then transferred through the VME bus. 
\end{enumerate}
Each board has 32 channels; in order to match the configuration of our
TPC, where the wires are grouped in bunches of 25, 2 bunches per
plane, we send the signal from 25 wires to each board, leaving the
remaining 7 channels floating.  

\subsection{Liquid Ar purification}

The necessity to drift free electrons in LAr over distances ranging
from tens of centimeters to several meters sets very stringent
requirements on the purity of LAr. For an applied electric field of
100V/cm the drift velocity of electrons in LAr is 0.5~mm/$\mu$s,
therefore an electron lifetime of several hundreds $\mu$s is required 
for a drift region with a depth of 10 cm. 
This corresponds to a contamination of O$_2$-equivalent impurities at
the level of few tens of parts per trillion (cf. few parts per million
in commercially available ultra-pure LAr).
The ICARUS collaboration has demonstrated that Ar can be purified
to this level in liquid phase, with high throughput
\cite{Cennini:LAr_pur}.  
The technology for the filters used to purify LAr for the Yale LAr TPC
has been developed at FNAL \cite{pordes}. A detailed paper is in
preparation. 
The filter is made of a copper alumina catalyst
\footnote{Engelhard Copper Alumina catalyst CU-0226S}, packaged in  a
CF nipple with sinterized steel caps; the filter used at Yale measures
55~cm long, with a diameter of 7~cm. Once exhausted, the
filters can be regenerated in house, heating them to 250$^{\circ}$ C
while flushing with a mixture of Ar gas and hydrogen (with a ratio of
Ar:H of 95:5). The filter used at Yale has been provided by the FNAL
group, and is routinely regenerated at Yale. 
Before introducing purified LAr the filling lines are kept under
vacuum, and the first ten liters of LAr passed through the filter are
dumped in a separate vessel, to reduce the risk of contamination.  
Commercial grade LAr, with an O$_2$ contamination of few ppm, is
purified with a single pass through the filter, at a rate of
$\sim$60~l/hr.  
A purity monitor \footnote{Built according to the ICARUS design
  \cite{t600} and provided by the FNAL group} was mounted underneath
  the TPC.
This has allowed quick and reliable measurements of the electron
  lifetime independently of the TPC operation. 
A loss of drifting charge due to attachment to impurities of less than
40\% over a drift time of 0.5~ms (equivalent to an electron lifetime
better than 1 ms) has been repeatedly measured in the TPC vessel,
stable over a period of 24 hours without recirculation.   
Using the same filter in two back-to-back experiments, without 
regeneration, we have noticed a drastic deterioration in the purity
during the second experiment. This may be an indication that the
filter itself is exhausted after purifying 500 to 700 l of commercial
Ar. The issue is currently being addressed in a more systematic way.   


\section{Results}

The LAr TPC has been tested on readily available cosmic rays. 
During data taking the electric field in the drift region was 100~
V/cm, and 300~V/cm between the induction and the collection planes. 
The data acquisition was triggered on the sum coincidence of hits on
one board (25 channels) of the collection plane.  
The drift velocity in LAr at 100~V/cm is $\sim$ 0.5~mm/$\mu$s, with a
maximum drift time of approximately 320~$\mu$s.  

\begin{figure}[htb]
\centering
  \includegraphics[width=.85\textwidth]{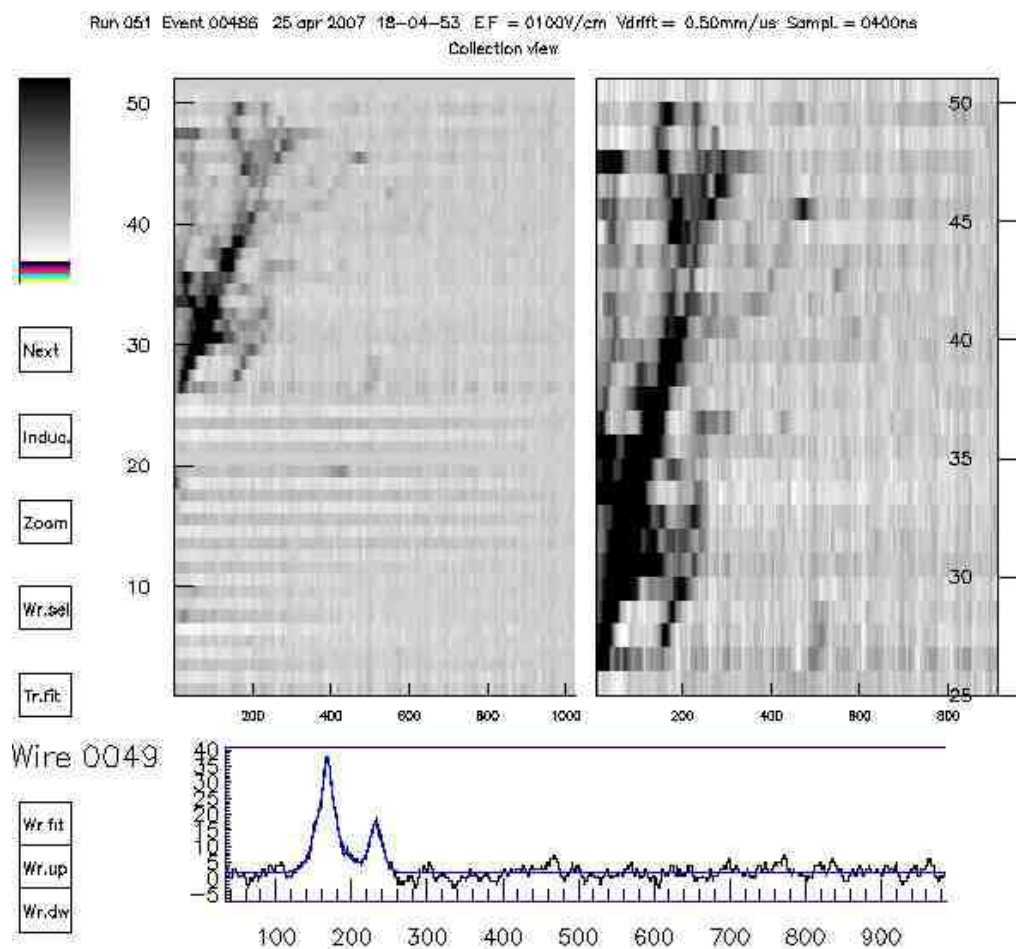}
  \caption{ Display of an electromagnetic shower, collection
  view. {\it Left:} full 2D image of the fiducial volume, shown as
  wire number {\it vs.} drift time (in units of 0.4~$\mu$s). The
  gray-scale is linear with the amplitude of the signal on each
  wire. {\it Right:} zoomed view, with contrast increased by a factor
  of two. {\it Bottom:} a digitized wire waveform, showing amplitude
  (in ADC counts.) {\it vs.} drift time, with two peaks (tracks)
  clearly identified. }    
\label{f2a}
\end{figure}

\begin{figure}
  \centering
  \includegraphics[width=.5\textwidth]{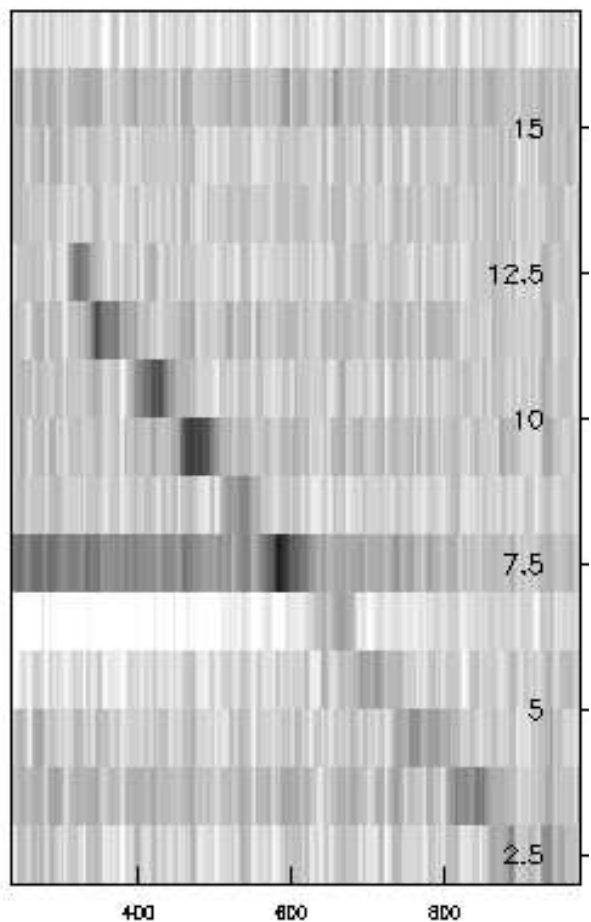}
  \caption{ Muon crossing the TPC (collection
    view). }
\label{f2b}
\end{figure}

\begin{figure}
  \includegraphics[width=.85\textwidth]{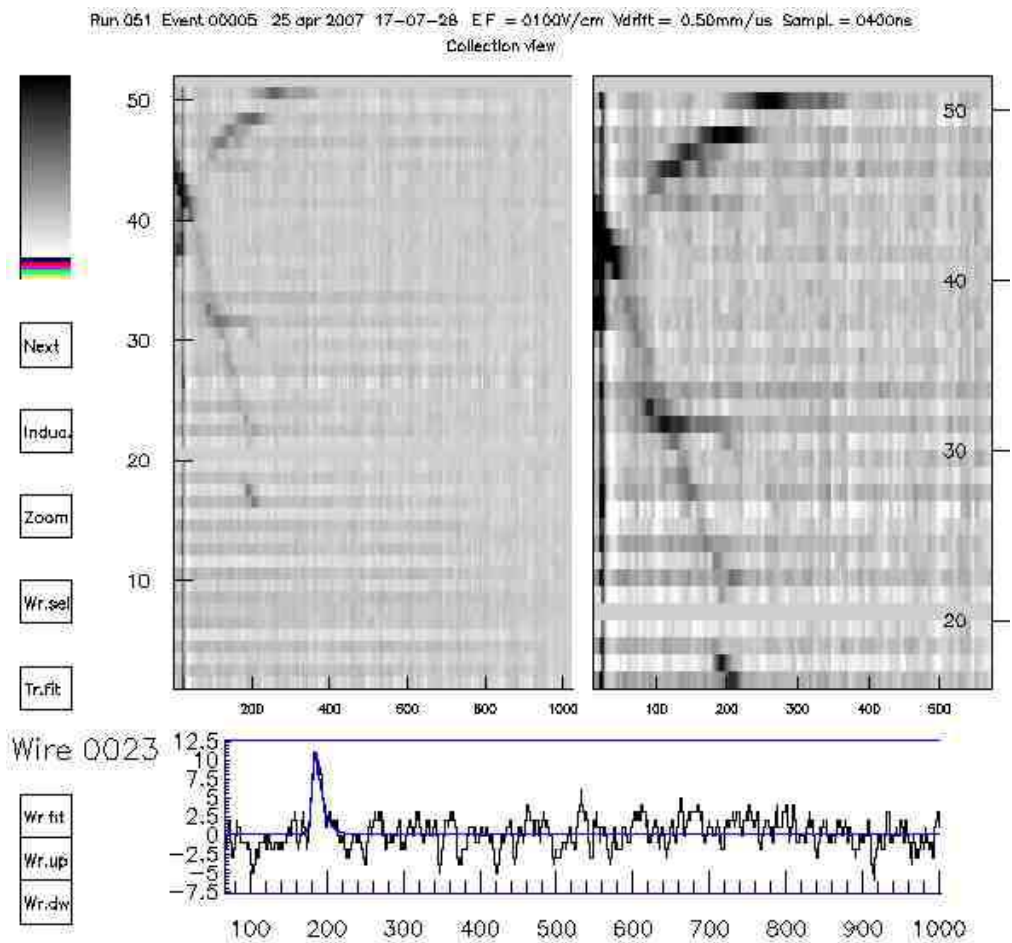}
  \caption{ Low multiplicity hadronic interaction (collection view). }
\label{f2c}
\end{figure}

\begin{figure}
  \centering
  \includegraphics[width=.45\textwidth]{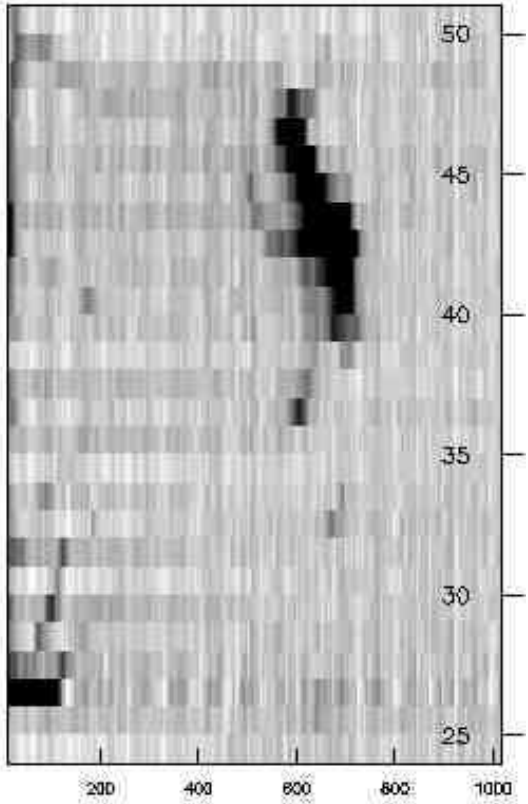}
  \caption{ Hadronic interaction with a blob of high density of
  deposited charge (collection view). }
\label{f2d}
\end{figure}

\begin{figure}
  \centering
  \includegraphics[width=.4\textwidth]{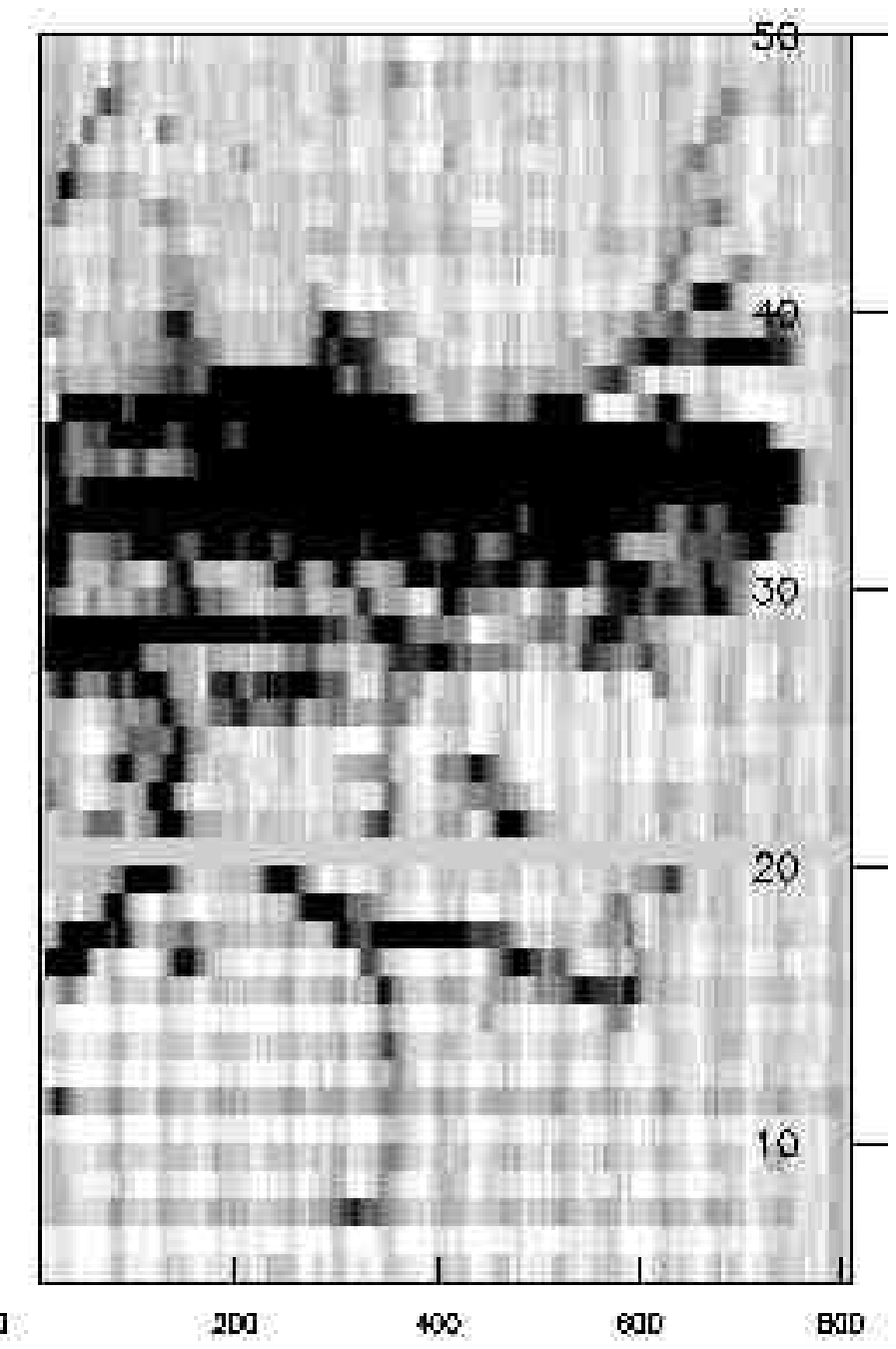} \\
  \includegraphics[width=.7\textwidth]{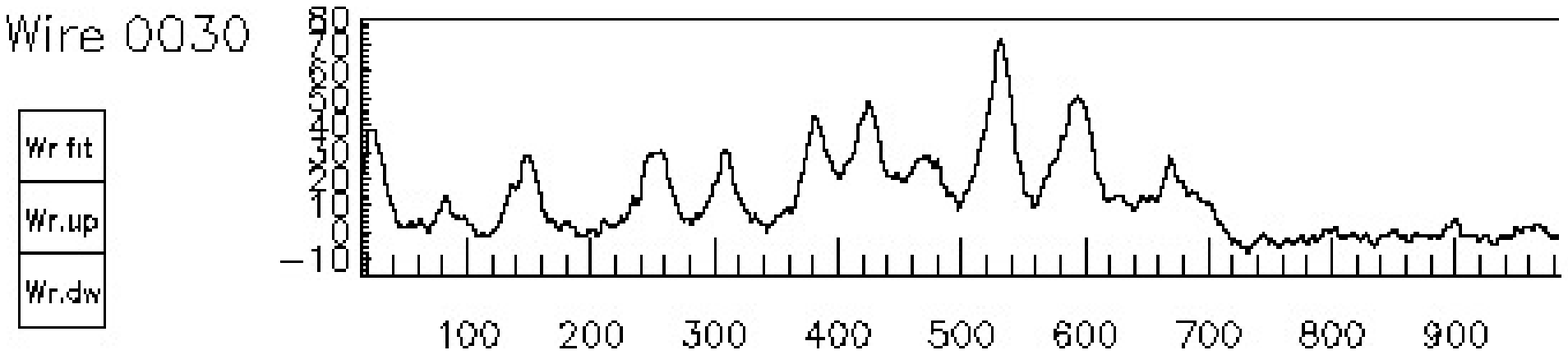}
  \caption{ An electromagnetic shower, which fills up the entire
  fiducial volume. Several individual tracks are visible in the wire
  waveform shown in the bottom window. }  
\label{f3}
\end{figure}

Some events recorded during a cosmic ray run are shown in
Fig.~\ref{f2a},~\ref{f2b},~\ref{f2c},~\ref{f2d} and \ref{f3}. 

\begin{enumerate}
\item {\it Fig.~\ref{f2a}:} Display  of an electromagnetic shower
  (collection view). In this and in all the figures shown here, the 2D
  view displays wire number {\it vs.} drift time (in units of
  0.4~$\mu$s). The gray-scale is linear with the amplitude of the
  signal on each wire. On the left is the full 2D image of the
  fiducial volume. On the right is the zoomed view, with contrast
  increased by a factor of two. At the bottom is a digitized wire
  waveform for one of the wires in the above plots. Two peaks (tracks)
  are clearly identified. The display show  the raw data, i.e. no
  filter has been applied offline.      
\item {\it Fig.~\ref{f2b}:} Image of a muon crossing the
  TPC (collection view). 
\item {\it Fig.~\ref{f2c}:} A low multiplicity hadronic interaction
  (display as in Fig.~\ref{f2a}).
\item {\it Fig.~\ref{f2d}:} Hadronic interaction with a high density
  of ionization (collection view); individual tracks are not
  resolved. 
\item {\it Fig.~\ref{f3}:} An electromagnetic shower that extends over 
  the entire fiducial volume; the core of the shower saturates the
  dynamic range of the electronics (collection view and single wire
  waveform with several tracks identified). 
\end{enumerate}

\begin{figure}
  \centering
  \includegraphics[width=.75\textwidth]{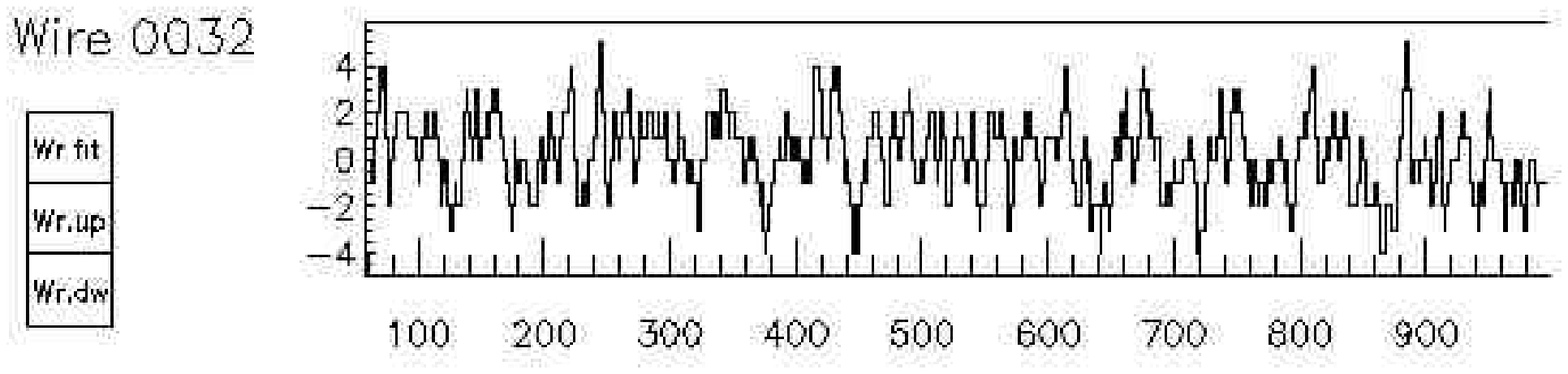} \\

  \includegraphics[width=.75\textwidth]{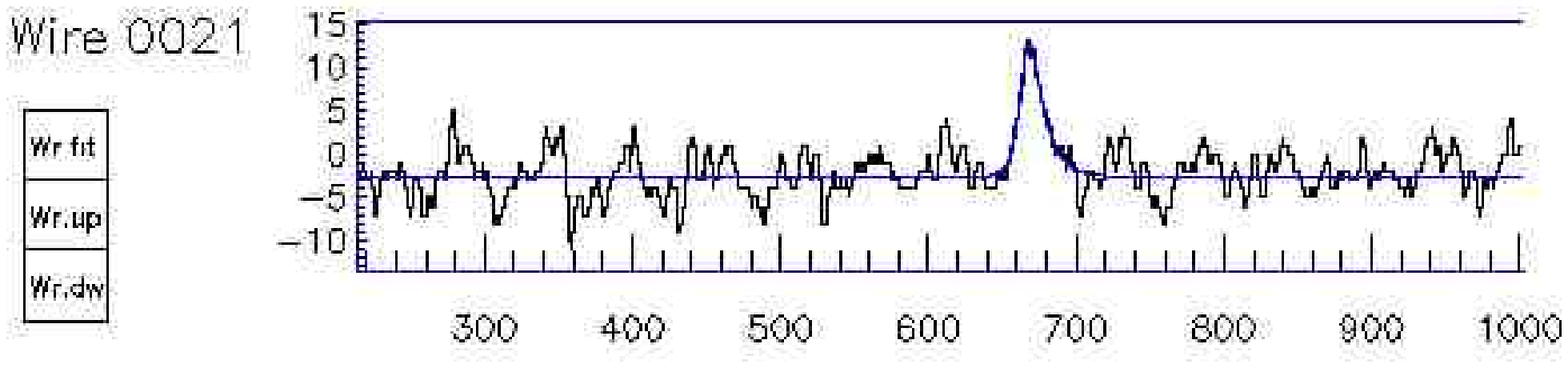}
  \caption{ {\it Top:} Waweform for an empty wire, in ADC counts {\it
  vs.} time [0.4~$\mu$s]. {\it Bottom:} signal induced by a minimum
  ionizing particle. }
\label{fwire}
\end{figure}

\begin{figure}
  \centering
  \includegraphics[width=.45\textwidth]{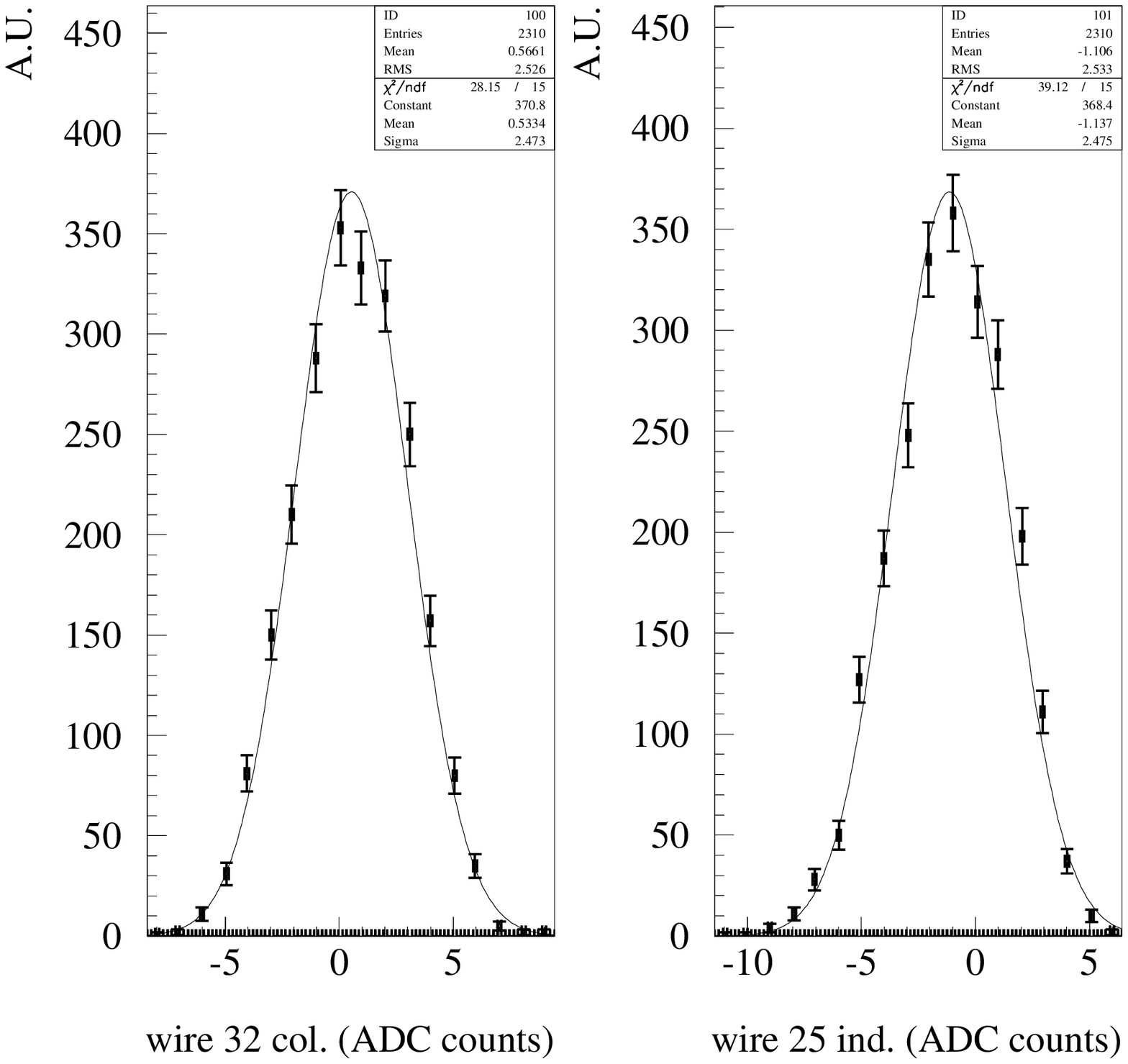} 
  \caption{ {\it Left:} distribution of the amplitude in ADC counts
  for an empty collection wire, i.e. the electronic noise. {\it
  Right:} the same for an induction wire. }
\label{fENC}
\end{figure}

The average noise is $\sim$2.5 ADC counts (RMS), giving a
signal-to-noise ratio of $\sim$4 for a minimum ionizing particle 
(mip) \footnote{This is more than a factor of two worse than the
signal-to-noise reported in \cite{t600} or in \cite{50l} (with
different front-end electronics); a large fraction of the noise in the
present setup is microphonic, due to difficulties in shielding the 
signal cables between the signal feedthrough and the front-end
electronics.}.   
Given the wire pitch of 5~mm and an energy loss of 2~MeV per cm
for a mip in LAr, the noise is roughly equivalent to 250 keV of energy
lost through ionization by a relativistic particle (Fig.~\ref{fwire}
and \ref{fENC}). Given a W-value of 23.6~eV in LAr, this corresponds
to $\sim$10,000 electrons produced in LAr by ionizing radiation.    


\begin{figure}
  \centering
  \includegraphics[width=.4\textwidth]{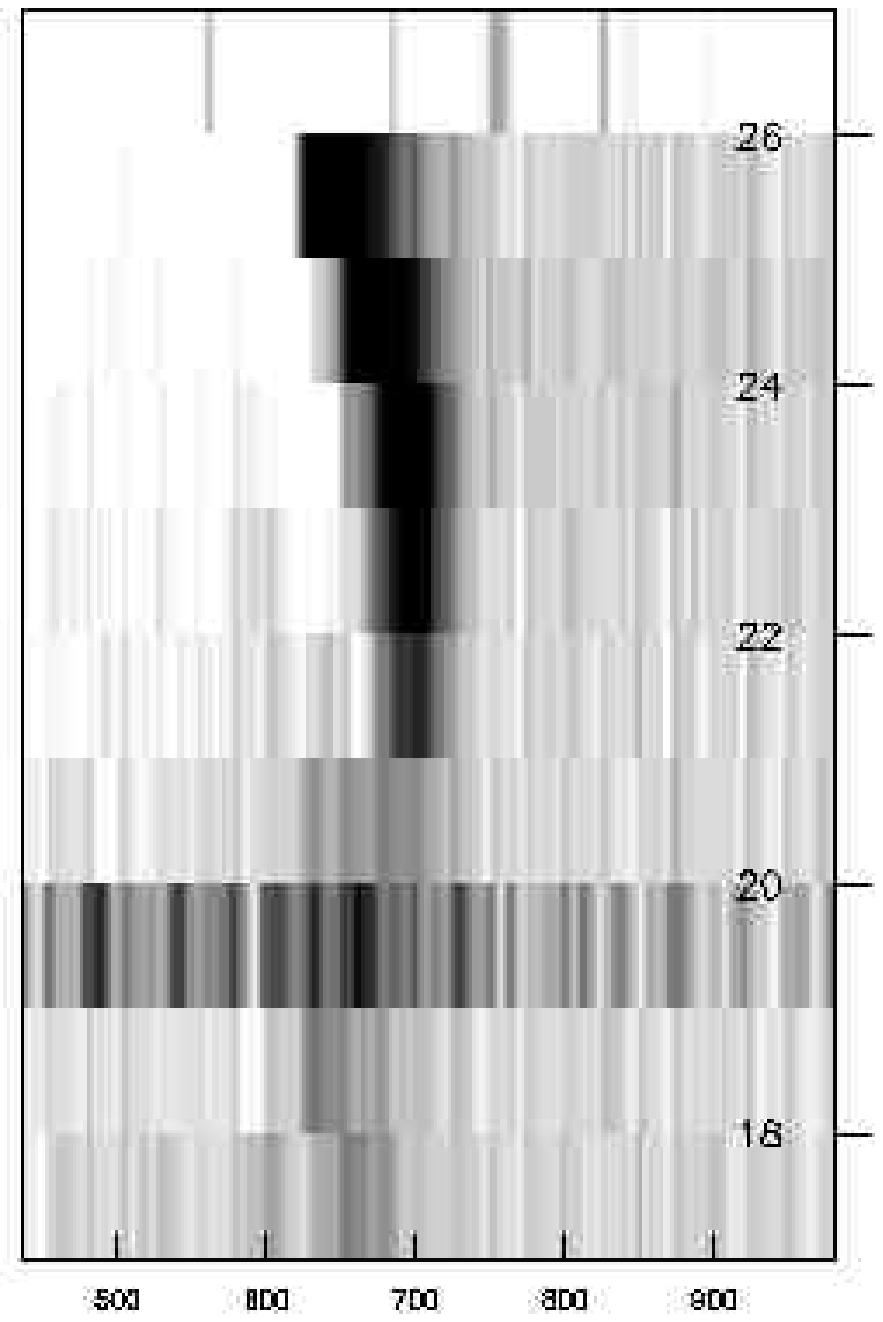} \\
  \includegraphics[width=.6\textwidth]{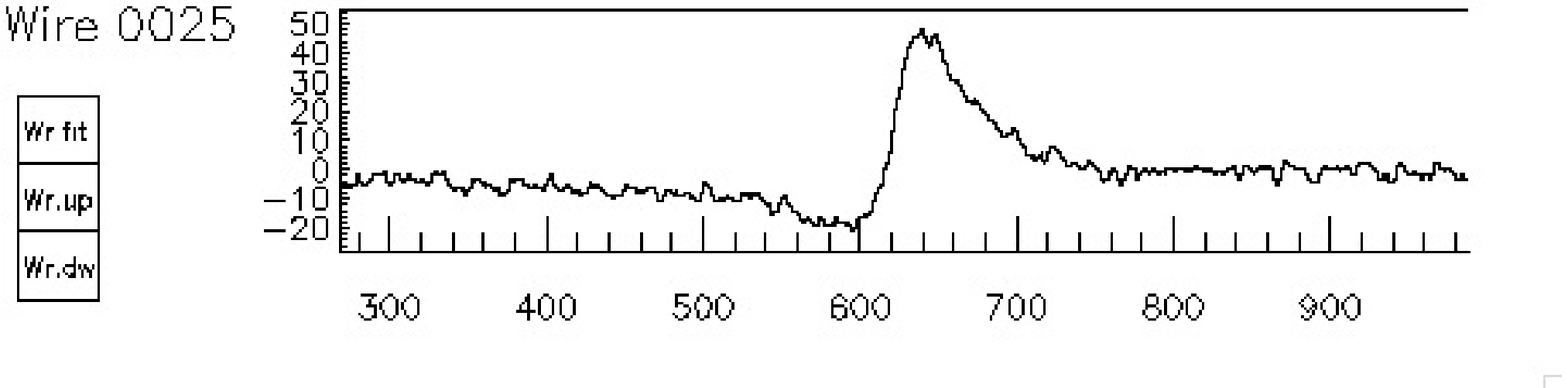}
  \caption{ {\it Top:} Induction view of the events shown in
  Fig.~\ref{f2d}, corresponding to the high density ``blob''. {\it
  Bottom:} waveform of an induction wire from the same event.}  
\label{ind}
\end{figure}

The signals from 50\% of the induction wires were 
noisy due to capacitive coupling with the high voltage in the drift
region, therefore the induction plane has not been used systematically
to provide full 2D images. Examples of partial 2D images and signal
waveforms are shown in Fig.~\ref{ind}. For the ``quiet'' wires, the
noise was comparable with the noise from the collection plane, as
shown in Fig.~\ref{fENC}.  

\section{Conclusions and outlook}

A prototype LAr TPC has been designed, built and tested at Yale over a
period of two years. It has been developed as a handy R\&D tool, with
the possibility of repeated runs over a short period of time. The LAr
purification is a recent development, applied for the first time to a
working imaging instrument, while  the readout electronics has been
provided by the ICARUS collaboration. 
The success in imaging cosmic rays marks an important milestone in
terms of technology transfer for the US LAr TPC effort.    

Steps are now being taken to expand upon this result. In addition to 
reducing the electronic noise in the system, the existing Yale TPC
will be augmented with a PMT to study the detection of scintillation
light in a LAr TPC. Other studies involving thick gaseous electron
multipliers (THGEM) \cite{THGEM} coupled with a LAr TPC are also being 
undertaken at Yale.   
The ArgoNeuT (Argon Neutrino Test) experiment \cite{argoneut} has been
designed and is currently being built, with the goal to begin running
in 2008 and to amass a large sample of neutrino interactions, about
200 per day, in the Fermilab Low Energy (LE) NUMI beamline.    
This test beam experiment, using front-end and DAQ electronics
designed and built by US institutions, will provide valuable
experience in operating a LAr TPC in a real beam environment, and a
large sample of low energy neutrino interactions. 


\ack{ We thank all the people who contributed to this project. In
particular, we acknowledge essential help from: S. Centro, S. Ventura,
B. Baibussinov and the ICARUS group at INFN Padova, for the 
readout electronics, software for DAQ and event display, and
round-the-clock support; N. Canci and F. Arneodo of LNGS, in the
initial work on LAr purification; the FNAL LArTPC group, in particular
H. Jostlein, C. Kendziora and S. Pordes, for providing the filters,
various equipment, and many excellent suggestions; M. Harrison,
S. Cahn and C. E. Anderson of Yale; L. Bartoszek of Bartoszek
Engineering.} 

\vspace{1cm}

\end{document}